\documentclass[journal,11pt,draftcls,onecolumn]{IEEEtran}
\usepackage{amsmath,amssymb,amsbsy,amsthm,geometry}
\usepackage{algorithm} 
\usepackage{algpseudocode} 

\let\emptyset\varnothing
\let\epsilon\varepsilon
\let\phi\varphi

\let\epsilon\varepsilon

\newtheorem*{lemma*}{Lemma}

\begin{document}


\title{A general method for the development of constrained codes} 

\author{
 \IEEEauthorblockN{Boris Ryabko\\}
 \IEEEauthorblockA{Federal Research Center for Information and Computational Technologies 
\\
Novosibirsk\\}}
\date{}

\maketitle


\begin{abstract}

Nowadays there are several classes of constrained codes intended for different applications.  The following two large classes can be distinguished. 
 The first class contains codes with local constraints; for example, the source data must be encoded by binary sequences containing no sub-words 00 and 111. 
 The second class contains codes with global constraints;  for example,  the code-words must be binary sequences of certain even length with half zeros and half ones.
  It is important to note that often the necessary codes must fulfill some requirements of both classes. 

In this paper we propose a general polynomial complexity method for constructing codes for both classes, as well as for combinations thereof. 
The proposed method uses the enumerative Cover's code, but the main difference between known applications of this code is that the known algorithms require the use of combinatorial formulae when applied, whereas the proposed method calculates all parameters on-the-fly using a polynomial complexity algorithm.

\end{abstract}

\section{Introduction} 
In modern transmission and storage systems, source information is converted using 
 data compression methods, self-correcting codes and constrained codes.
The purpose of data compression and correction coding is to reduce the amount of data before transmission and to include additional data to correct the information after transmission or storage, respectively. The purpose of constrained encoding is to transform the original data before transmission to avoid all (or most) errors. Of course, all the codes mentioned above can be used together to improve performance.

Constrained codes have been intensively researched and applied in practice since the middle of the 20th century, when various hard and optical discs became widespread \cite{imm-rev}.  Nowadays, the constrained codes are used in many kinds of  storage devices and numerous  data transmission systems  \cite{imm-rev,imm-book,zig,cold1} and these applications are based on profound theoretical results developed in numerous publications, see
 \cite{imm-rev,imm-book,cold1} for reviews.  

The so-called runlength-limited  codes (RLL) were perhaps 
 the first class of constrained codes. In general, this class can be defined as follows: there exists a set  of forbidden words  (or patterns) $U= \{u_1, u_2, ... \} $ and any codeword must not contain any forbidden subwords  \cite{imm-book,zig,cold1}. For example, if $u_1 = 00, u_2 = 111$, the codeword set of the three-letter constrained code is as follows: $010, 011,101,110$ (hence, this code can be used to encode 2-bit source words). 
   Various RLL codes have been developed for many sets of forbidden patterns designed to meet the requirements of different storage devices and transmission systems, and there are now simply realisable and efficient algorithms for many RLL codes, see \cite{imm-book,zig,cold1} for a review.

Another large class of the constrained codes   requires some constraints for the codeword as a whole. To describe several such codes, it will be convenient to represent the codewords in the alphabet $\{-1,+1\}$.
So, the so-called balanced code is probably the first code of this type proposed by Gorog \cite{gor}.
If we denote the codewords of a balanced code  (BC) by $x= x_1 ... x_n$, $n$ is even, then by definition $x_1 +x_2 +.... x_n = 0$. Sometimes  a more general problem is considered where the sum is not zero but is bounded by some (small) number $\alpha$, that is, $|x_1 +...+x_n| \le \alpha$. 
 Later these codes were investigated by many researchers, and some  results   can be found in  \cite{imm-book,zig,Alon,kn,imm3,imm4} for general alphabets.

Another object of study is the limited current sum (LRS) code. By definition, for any of its codewords $x= x_1 ... x_n$ the current sum must satisfy 
 $|x_1 +x_2 + ... x_k| \le \delta$ for any $k = 1, 2, ... , n$, where $\delta > 0$ is a parameter. 
Note that for a codeword $x = x_1 ... x_n$ and any $k,l, k \le l$, we have 
 $|x_k +x_{k+1} + . . . . +x_l | \le 2 \delta$. So, the sum is limited for any so-called ``sliding window'' $x_k  . . . . x_l$.
For small $\delta$ and large $n$, LRS codes have a small zero frequency (dc) in the spectrum, and this property is very desirable for many storage devices and communication systems. This is why LRS codes have been developed by many researchers, see \cite{imm-rev,imm-book,zig} for a review.

We combain BC and LRS codes in one class by the following definition of the set of words $x_1 ... x_n$ from the alphabet $A$:
\begin{equation}\label{set} 
T_n(\delta_1,\delta_2,\alpha,\beta) = \{ x_1... x_n :     \delta_1 \le \sum_{i=1}^k x_i \le \delta_2    \,  for \, all \, k \,,  \, and \,\,
\alpha \le \sum_{i=1}^n x_i \le \beta \} \, .
\end{equation}
Indeed, for $\alpha = \delta_1, \beta = \delta_2$ ws obtain LRS constrained codes 
and for $ \delta_1 = n \, ( \min_{a \in A} a ) \,, 
 \delta_2= n\, (\max_{a \in A} a) $ we obtain BC constrained codes. 

The fourth class of codes with constraints is the so-called codes with energy constraints \cite{ener}. In this case, it is convenient to use the alphabet $\{0,1\}$ and it is assumed that any $1$ carries a unit of energy. (So, a sequence of letters from $\{0,1\}$ conveys not only information but also energy.)  In this problem, the density of units (i.e., the rate of energy transferred) must be bounded within some limits.

In this paper, we consider the following formal model: there exists a source of sequences in the alphabet $A$ consisting of integers (of type $\{0,1\}$ or $\{ -1,+1 \}$) and these sequences of some length $n$ must be encoded by a certain constrained code, that is, any source sequence $x$ encoded by a codeword $c(x) $ of a constrained code of definite length $m$ such that $c(x^1) \neq c(x^ 2)$ if $x^1\neq x^2$. If we denote the set of all possible codewords by $C$, it is obvious that $|C| \ge 2^n$. (Here and below $|u|$ is the length of $u$ if $u$ is a word, and the number of elements if $u$ is a set.)

In 1973, Cover proposed the so-called enumerative source coding \cite{cover}, which has been widely used for constrained coding \cite{imm-book,cold1,loco,kur}. It is worth noting that this enumerative coding was used prior to Cover's paper for the case of encoding $0-1$ words of a certain length $n$ with a fixed number $m$ of units (and obviously $n-m$ zeros) \cite{lynch,dav,babkin}.

Generally speaking, Cover's code can be used to create any constrained code, but such applications require some a priori combinatorial analysis.  For example, the code for the mentioned problem about $0-1$ sequences with a fixed number of units is based on binomial coefficients and Pascal's triangle, whereas for other constrained codes the combinatorial analysis and the obtained combinatorial formulas are more complicated {\cite{cold1,kur}. 

Results are now available for some pairs of constraints, especially combining RLL with some others, and in all cases the resulting codes are based on rather complex elaborated combinatorial formulas \cite{cold1,kur,a1,a2}.
Apparently, the absence of known combinatorial formulas and methods of their development is a significant obstacle for construction of new constrained codes for various problems.

In this paper we propose a new approach to constructing codes with constraints in which the required parameters are computed in polynomial time and no special combinatorial analysis is required.
The following is an example of a ``complex'' problem which can be solved by the suggested method without combinatorial analysis. We need to construct a constrained code for an $n$-letter sequence $x_1.... x_n$ of $\{0,1\}$ for which i) the density of units is at least $1/2$, ii) the sum of units does not exceed $\lfloor 	2n/3\rfloor 	$, iii) $ |\sum_{i=1}^k x_i - k/2| \le 20$ for $k=1,..., n$ and iv) the words $0011$ and $01010$ should be excluded. (Thus, this problem includes all four constraints mentioned above.)

The rest of the paper consists of the following. Parts 2 contains the description of the Cover's method and  the   limited current sum (LRS) code, the part 3 contains description  of codes with constraints for three other problems described above. The fourth  part describes codes for  combinations of constraints described above and studied in parts 2-3. In a small conclusion we will talk about complexity of the proposed algorithms,  some simplifications and generalisations of the described codes.

\section{Cover's algorithm}

In \cite{cover}   Cover suggested the following  general method of an  enumerative encoding. 
There is an $m$-letter alphabet $A = \{a_1, a_2, ..., a_m\}$ 
and let $A^n$ be  a set of of 
words of length $n$ over $ A$.    Every subset $S \subset A^n$
is called a source.      An enumerative code $f$ is given by two mappings $f_c : S \to
\{0, 1\}^l $, where $l= \lceil \log |S| \rceil $ and $f_d:  f_c(S) \to S$ , so that $f_d(f_c(s)) = s $ for
all $s \in  S$. The map $f_c$  is called an  encoder and $f_d$ is called
a decoder.  It is assumed that the alphabet $A$ consists of numbers. 
Let us describe an enumerative code from \cite{cover}. 

{ \it  \textbf{ Encoder:}}
Let $ N(x_1...x_k )$ be
the number of words which belong to $S$ and have the prefix 
$x_1 ... x_k, \, k=1, 2, ... , n-1$.
For $x_1x_2...x_n \in S $ define 
the code word   $f_d(x_1...x_n)$ by
\begin{equation}\label{N}
code
(x_1...x_n) =
    \sum_{i=1}^n        \sum_{a < x_i}
N(x_1 ... x_{i-1} a ) \, .
\end{equation}

{ \it  \textbf{ Decoder:}}
Let us describe the decoder.   Denote $\alpha = code
(x_1...x_n)$, and  $b_{t+1}  = N( )+1$, where $N( )=   \sum_{a \in A}
N( a ) \, (= |S|) $. 
Then calculates  $b_1 = N(a_1), b_2= N(a_2), ... , b_m=N(a_m)$. If $  b_j \le \alpha <
b_{j+1}$ then the first letter $x_1$ is  $ a_j$.  In order to find $x_2$  the algorithm calculates 
$\alpha = \alpha - N(a_j), b_{m+1} = N(a_j)+1,  b_j = N(x_1 a_j),  j =1, ..., m$. 
If $  b_k \le \alpha <
b_{k+1}$ then the second  letter $x_2$ is  $ a_k$. 
Then, $x_2 = a_k$ and so on.

Note that the complexity of Cover's method is determined by the complexity of computing $ N(\,), $ and hence, developing a simple method for computing $ N(\,)$ is the main problem. Currently, the use of Cover's method is based on combinatorial formulas to compute $N(\,)$, as shown in the following example of encoding binary words of $n$-letters with a given number of units $\nu$. That is, the alphabet $A=\{0,1\}$, set $S$ contains words for which $\sum_{k=1}^n x_i = \nu$.
In this case
\begin{equation}\label{bin} 
  N(x_1...x_{k-1}0) = \binom{n-k}{\nu - \sum_{i=1}^{k-1}   x_i      }  \,.
\end{equation}
Using this formula and (\ref{N}) we obtain
$$  code(x_1 ... x_n) = \sum_{k=1}^n  \binom{n-k}{\nu - \sum_{i=1}^{k-1}   x_i      } \, .
$$
(This solution was found before Cover's code was described, see \cite{lynch,dav,babkin}).
The set of binomial coefficients (\ref{bin}) can be stored in memory or computed on the fly as needed.

Nowadays combinatorial formulas for many codes with constraints are known and widely used in practice.
But sometimes such formulas can be very complicated, and for some interesting problems these formulas are not developed at all.

In this paper we propose a direct computation of the sums $N(\,)$,
 not based on any combinatorial formulae. These $N(\,)$ can then be stored in memory or computed on the fly. More precisely, we describe polynomial methods for computing the values of $N( \,)$ for the various constrained codes discussed in the introduction.

\section{Polynomial complexity method for BC and  LRS codes }


In this part we consider the following set $T_n(\delta_1,\delta_2, \alpha, \beta)$, $ \,\, \delta_1 \le  \alpha \le  \beta \le \delta_2, $ of sequences $x_1 ... x_n$, $x_i $ from some alphabet $A$,
see (\ref{set}). 
It is worth noting that the size of this set  grows exponentially.
For example,  for $A= \{-1,1\}$,  $|T_n(-1,1,-1,1)| = 2^{\lfloor n/2 \rfloor } $. So,
our goal is to develop a simple 
 method for calculating $N(x_1...x_r )$ for $r= 1, ... , n-1$ for  $x_1 ... x_n  \in $ $T_n(\delta_1,\delta_2, \alpha, \beta)$, because this is a key part of the Cover encoding and decoding method (described in the previous section).

First, let us give some comments to make the description of the algorithm clear. 
The key observation is the following: the number of trajectories with the prefix $x_1...x_k$ depends only on the sum $x_1+...+x_k$ and hence the algorithm does not need to store the  word $x_1...x_k$.  This 
 reduces the required memory from exponential to polynomial.
The algorithm  computes 
 the number of trajectories (or words) that start with $z_r = \sum_{i=1}^r x_i $.  
The table 
 $S[i,j], i= 1,...,n-r, j= \delta_1, ..., \delta_2$ contains the numbers of trajectories of type $z_rx_{r+1} ... x_{r+i}$, $i= n-r$.
In general, when the algorithm goes from $r+i$ to $r+(i+1)$, we must extend all current trajectories by letters 
 $a \in A$, and the table $S[,]$ contains the number 
  of trajectories.

\begin{algorithm}[H]
\begin{algorithmic}[1]
\State \textbf{Input:} $x_1 \ldots x_r$
\State \textbf{Output:} $N(x_1 \ldots x_r)$
\State Create table $S[i,j]$, $i=  1, \ldots, n-r+1$,  $j=\delta_1, \ldots,  \delta_2$
\State Fill $S$ with zeros
\State $z_r \gets \sum_{i=1}^r x_i$
\State $S[1,z_r] \gets 1$
\For {$i \gets 1$ to $n-r$}
    \For {$k \gets \delta_1$ to $\delta_2$}
        \For {$a \in A$}
\If {$ \delta_1 \le k + a \le \delta_2 $}
State $S[i+1, k+a]  \gets   S[i+1, k+a]+ S[i,k]$ 
\EndIf

\EndFor
\EndFor
\EndFor
\State {$N \gets 0$}
\For {$i \gets \alpha$ to $\beta$}
\State $N \gets N+S[n-r+1,i]  $
\EndFor

\end{algorithmic}
\caption{Computation of $N$}\label{alg:N}
\end{algorithm}

Let us consider a small example.
Let  $A= \{-1, +1 \}$,  $n=6$,  $x\in $ $T_n(\delta_1,\delta_2, \alpha, \beta)$,  where $\delta_1 =0, \delta_2 = 3, \alpha = 0, \beta=2$ and  let  the algorithm  1  be applied for calculation $N(+1\,-1\, +1)$. 
Then $ z_3=1$, and the algorithm is carried out as follows:
$$
S[1,\,] = 
 \begin{pmatrix}
   0 \\
  0\\
1\\
  0
 \end{pmatrix}
S[2,\,] = 
 \begin{pmatrix}
   0 \\
  1\\
0\\
  1
 \end{pmatrix}
S[3,\,] = 
 \begin{pmatrix}
   1\\
  0\\
2\\
  0
 \end{pmatrix}
S[4,\,] = 
 \begin{pmatrix}
   0\\
  3\\
0\\
  2
 \end{pmatrix}
$$
Applying the  last algorithm cycle from $\alpha =0$ till $\beta=2$ we obtain $N=2+0+3=5$ and, hence,  
$N(+1\,-1\, +1) = 5$.


Now let us estimate the complexity of the described method. 
During decoding and encoding the values $N(x_1 ... x_i)$ are calculated and for any $N$ 
the summation $O(n )$ is required. So, the total time (for $i=1,..., n-1$) is $O(  n^2)$. 
The required memory size is $O(n^2)$, but   it is sufficient to store only two columns of the table 
 $S$, and hence the required memory is $O(n)$. Taking into account that $N(x_1...x_k)$ depends only on $\sum_{i=1}^k x_i$, we see that there are only $(\delta_2 - \delta_1)$ of different $N$s and all of them can be calculated in advance and stored in $O((\delta_2 - \delta_1)n)$ cells.   



\section{Energy constraints} 

In \cite{ener} the so-called subblock energy-constrained codes (SECC) and sliding window codes (SWCC) are considered for the $\{0,1\}$ alphabet.
To describe them, it will be convenient to call the number of units in the word $u$ with weight $u$ and denote it by $w(u)$.

So-called sub-block energy-constrained codes (SECC) are binary sequences of length $n=ml$, which are treated as a sequence of $m$ $l$-bit sub-blocks ($m$ and $l$ are integers). The weight
of each subblock is at least $\alpha$ and at most $\beta$, where $\alpha$ and $\beta,  \alpha <\beta$ are some integers.  This set of sequences equals $T_n(0,  l,  \alpha, \beta)$ (see (\ref{set}) ) and hence the algorithm presented in the  previous part is applicable. 

Sliding window constrained codes are  $n$-bit words $y_1 ... y_n$ such that the weight of any $l$-bit subword $ x_{i+1} ... x_{i+l}$ is equal to minimum $\alpha$ and maximum $\beta$, i.e.  $\alpha \le w(x_{i+1} ... x_{i+l}) \le \beta$
 (here $i$ can be any integer from $[1,n-l]$, not necessarily a multiple of $l$).

So,  the following problem of encoding is considered.     There is a set of sequences $x_1, ... , x_n$, $x_i \in \{0,1\} $ and $ \alpha \le \sum_{i =k}^{k+l-1} x_i $ $\le \beta$ for any integer $k \in [1,...,n-l+1] $     
and let denote the set of such words by  $SWCC_n(\alpha, \beta)$.   

\textbf{The algorithm for sliding window-constrained codes.}    
To simplify the description of the algorithm, we will use an additional letter $\emptyset$ and hence the alphabet will be $\{\emptyset,0,1 \}$. We also assume that the weight of $\emptyset$ is 0.
The algorithm uses the table $S[i,u]$, where $i$ is an integer, $i= 1, ..., n-r+1$ and $u$ is a  word,
$u \in \{ \emptyset,0,1\}^l$. (This table will be used to store the number of trajectories.)
Also we denote 
 $\nu(v)$ the number of the letters $\emptyset$ in the word $v$.
Note that we use the letter $\emptyset$ to simplify the description of the algorithm. In fact, this letter is needed only for the case $r< l$.

\begin{algorithm}[H]
\begin{algorithmic}[1]
\State \textbf{Input:} $x_1 \ldots x_r$. Comment: $x_1 \ldots x_r \in SWCC_r(\alpha, \beta)$. Otherwise, $N=0$. 
\State \textbf{Output:} $N(x_1 \ldots x_r)$
\For {$i\gets -l+1$ to $0$ }
\State $x_i \gets \emptyset$ 
\EndFor
\State Create table $S[i,u]$, $i=  1, \ldots, n-r+1$,  $u \in \{\emptyset,0,1\}^l$
\State Fill $S$ with zeros
\State $S[1,x_{r-l+1} x_{r-l+2}  \ldots x_r] \gets 1$
\For {$i \gets 1$ to $n-r$}
 \For {$u \in \{\emptyset,0,1\}^l$}
\State $v0 \gets u_2 u_3 \ldots,  u_{l-1} 0, \,\,\, v1 \gets  \, u_2 u_3 \ldots,  u_{l-1} 1$

\If {$ w(v0) \ge \alpha - \nu(v0)$} 
\State
 $S[i+1, v0]  \gets   S[i+1, v0]+ S[i,u]$ 
\EndIf
\If {$ w(v1) \le \beta $} 
\State
 $S[i+1, v1]  \gets   S[i+1, v1]+ S[i,u]$ 
\EndIf
      \EndFor
\EndFor
\State $N \gets 0$
\For {$u \in \{0,1\}^l $}
State $N \gets N+ S[n-r,u] $
\EndFor

\end{algorithmic}
\caption{Computation of $N$}\label{alg:N}
\end{algorithm}

Let us estimate the complexity of this algorithm. 
Table $S[,]$ contains $3^l (n+l)$ cells and the computation time is proportional to the same value. 
It is clear that this algorithm can be significantly simplified if $\emptyset$ is removed, but the asymptotic complexity estimate will be the same, so we will not describe these simplifications.

\section{The method for RLL codes }
Let codewords of a RLL code be $n$-letter words over an alphabet  $A= a_1, ... , a_m,  m\ge 2$, and  $V= \{v_1, ..., v_s\}, s \ge 1,$ be the set of forbidden  words, that is, any codeword $y_1 ... y_n$   does not contain any $v \in V$ as a subword. Denote the set of such words by $RLL_n(V)$. 
As before, we will use Cover's method, and the description of the algorithm will be reduced to computing the numbers $N(x_1... x_r)$, see the description in part 2.  It is worth noting that the described algorithm will be very close to the algorithm for sliding window codes. In particular, we extend the alphabet $A= \{ a_1, ... , a_m \}$ to $ A' = \{\emptyset,a_1, ... , a_m \}$.

\textbf{The algorithm for RLL codes.}    
Let us define 
\begin{equation}\label{V}
l_v = |v|,  \mu= \max_{v \in V} l_v.
\end{equation}

The algorithm uses the table $S[i,u]$, where $i$ is an integer, $i=0, 1, ..., n-r+1$ and $u$ is a  word,
$u \in A'^\mu$. 

\begin{algorithm}[H]
\begin{algorithmic}[1]
\State \textbf{Input:} $x_1 \ldots x_r$. Comment: $x_1 \ldots x_r $ does not contain subwords from $V$ . Otherwise, $N=0$. 
\State \textbf{Output:} $N(x_1 \ldots x_r)$
\For {$i\gets -\mu+1$ to $0$ }
\State $x_i \gets \emptyset$ 
\EndFor
\State Create table $S[i,u]$, $i=  1, \ldots, n-r+1$,  $u \in A'^{\,\mu}$
\State Fill $S$ with zeros
\State $S[1,x_{r-\mu+1} x_{r-\mu+2}  \ldots x_r] \gets 1$

\For {$i \gets 1$ to $n-r$}
 \For {$u \in  A'^{\,\mu}$} 
\For {$a \in A'$} 
\State$\lambda \gets \textbf{true}$,   
$w \gets  u_2u_3 ... u_\mu a$
\For {$v \in V$}
\If {$ w_{\mu-l_v+1} ... w_\mu = v$}  \State $\lambda \gets \textbf{felse} $,  \EndIf
      \EndFor

\If {$ \lambda =  \textbf{true} $} 
 \State $  S[i+1, w] \gets S[i+1, w] + S[i,u]$

 \EndIf
\EndFor
\EndFor
\EndFor
\State $N \gets 0$
\For {$u \in A^\mu $}
State $N \gets N+ S[n-r+1,u] $
\EndFor

\end{algorithmic}
\caption{Computation of $N$}\label{alg:N}
\end{algorithm}

\section{Constructor for restricted codes}

Quite often codes satisfying various constraints are investigated in connection with their application to certain data transmission and storage systems. Generally speaking, sometimes such problems can be solved by performing special combinatorial studies to find expressions for $N(\,)$ in Cover's method, see, for example, \cite{imm-book,cold1,loco,kur,a1,a2}.  
In this part, we show that the described methods of directly computing the values of $N(\,)$ can be easily used in the case of joint application of multiple constraints.  

As an example,   we consider the case where constraints on the sums of sequence letters and RLL constraints are given together. All other possible combinations of constraints (e.g., energy and RLL) are similar.  So we consider the following set $T_n(\delta_1,\delta_2, \alpha, \beta)$, $ \,\, \delta_1 \le  \alpha \le  \beta \le \delta_2,$ of sequences $x_1 ... x_n$, $x_i $ from some alphabet $A$,
see (\ref{set}).  Also there is a   set with RLL constrains, that is a set of forbidden  words
  $V= \{v_1, ..., v_s\}, s \ge 1,$  and any codeword $x_1 ... x_n$   does not contain any $v \in V$ as a subword.  (This set was denoted by $RLL_n(V))$. 
So our goal is to  build the constrained code for  the set of sequences $\Delta = T_n(\delta_1,\delta_2, \alpha, \beta)$ $\cap \, RLL_n(V)$. 
As before, we will use Cover's method and therefore only describe the computation of the numbers $N(x_1... x_r)$.
To do this, we combine Algorithm 1 and Algorithm 3 as follows.

The algorithm is designed to compute the number of trajectories (or words) that 
 belong to $\Delta$.
We consider the table $S[i,j,u]$, where $i= 1,...,n-r, j= \delta_1, ..., \delta_2$,   $u \in A'^\mu$,  ($\mu$ defined in (\ref{V}),) $A' = \{\emptyset,a_1, ... , a_m \}$.

\begin{algorithm}[H]
\begin{algorithmic}[1]
\State \textbf{Input:} $x_1 \ldots x_r$,  \textbf{Output:} $N(x_1 \ldots x_r)$
\State Create table $S[i,j,u]$, $i=  1, \ldots, n-r$,  $j=\delta_1, \ldots,  \delta_2$, $u \in A'^\mu$
\State Fill $S$ with zeros
\For {$i \gets -\mu+1$ to $0$ }
\State $x_i \gets \emptyset$ 
\EndFor
\State $z_r \gets \sum_{i=1}^r x_i$,  $\,\,S[1,z_r,x_{r-\mu+1} x_{r-\mu+2}  \ldots x_r] \gets 1$

\For {$i \gets 1$ to $n-r+1$}
    \For {$j \gets \delta_1$ to $\delta_2$}
        \For {$a \in A'$}

\If {$ \delta_1 \le j + a \le \delta_2 $} 
\State       $\lambda_1 \gets \textbf{true}$     
\Else
\State          $\lambda_1 \gets \textbf{false}$  
\EndIf
\State$\lambda_2 \gets \textbf{true}$,   
  $\,\,w \gets  u_2u_3 ... u_\mu a$
\For {$v \in V$}
\If {$ w_{\mu-l_v+1} ... w_\mu = v$}  \State $\lambda_2 \gets \textbf{felse} $ \EndIf
      \EndFor

\If {$ \lambda _1 \& \lambda_2 =  \textbf{true} $} 
 \State $  S[i+1, j+a,w] \gets S[i+1, j+a, w] + S[i,j,u]$

 \EndIf

\EndFor
\EndFor
\EndFor
\State {$N \gets 0$}
\For {$j\gets \alpha$ to $\beta$} \For {$u \in A^\mu$}
\State $N \gets N+S[n-r+1,j,u]  $
\EndFor
\EndFor

\end{algorithmic}
\caption{Computation of $N$}\label{alg:N}
\end{algorithm}

\section{Complexity of the constrained  codes}
In this part we will briefly evaluate the complexity of the developed codes. First of all, we note that there are many obvious simplifications of the described methods. Thus, in all algorithms the table $S[i,j]$ can contain only two rows corresponding to $i$, the third and fourth algorithms can be described without an extra letter $\emptyset$, etc. This was done to avoid unnecessary small details and hence make the description shorter and clearer.

Next, we briefly discuss asymptotic estimates of the memory and time required for the developed algorithms (\ref{N}). First of all, we note that Cover's method requires $O(n)$ summation operations with $O(n)$ integers of $O(n)$-bit length (because, as we showed in the introduction, the number of trajectories grows exponentially even in the simplest cases).  So, the running time and memory size  of Cover's algorithm is $O(n^2)$.

Now let us evaluate the complexity of computing the values of $N(\,)$ used in the algorithms described above, but first note that the following two ways of using these values are possible: either the set of values $N(\,)$ can be calculated in advance, stored and used several times for encoding and decoding different $x_1 ... x_n$, or $N(\,)$ can be computed again for any $x_1 ... x_n$. The upper estimate can be obtained by assuming that $\delta_1$ and $\delta_2$ in the algorithms are equal to $O(n)$. In this case, the computation time of $N(\,)$ values is $O(n^3)$ and the memory size is $O(n^2)$ bits, assuming that the length of $N(\,)$ is $ O(n)$ bits. 

The developed algorithms are not symmetric in the following sense: If, for example, $N(x_1 ... x_n)$ is computed, then in all described algorithms the length of summands increases from a few bits to $O(n)$ bits. The work \cite{fast} describes an algorithm for the Cover's method, where most of the operations are performed on numbers of the same length, which significantly reduces the complexity. It seems natural to reorganise the algorithms developed above in a similar way in order to reduce the complexity, but for this purpose a special new algorithm must be developed.

\end{document}